\documentclass{aastex}
\usepackage{spr-astr-addons}
\usepackage{url}

\RequirePackage{color}

\begin{document}

\title{Magnetized Particle Capture Cross
Section for Braneworld Black Hole}
\shorttitle{Magnetized Particles in Braneworld Model}
\shortauthors{Rahimov et al.}

\author{O.G. Rahimov\altaffilmark{}}
\author{A.A. Abdujabbarov\altaffilmark{}} \email{ahmadjon@astrin.uz} \and
\author{B.J. Ahmedov\altaffilmark{}}

\altaffiltext{1}{Institute of Nuclear Physics,
        Ulughbek, Tashkent 100214, Uzbekistan}
\altaffiltext{2}{Ulugh Begh Astronomical Institute,
Astronomicheskaya 33, Tashkent 100052, Uzbekistan}

\begin{abstract}

Capture cross section of magnetized particle (with nonzero
magnetic moment) by braneworld black hole in uniform magnetic
field is considered. The magnetic moment of particle was chosen as
it was done by \citet{rs99} and for the simplicity particle with
zero electric charge is chosen. It is shown that the spin of
particle as well as the brane parameter are to sustain the
stability of particles circularly orbiting around the black hole
in braneworld i.e.  spin of particles and brane parameter try to
prevent the capture by black hole.

\end{abstract}

\section{Introduction}\label{intro}

The study of the test particle motion and fields attracts more
attention of the astrophysicists  because they play an important
role in the study of astrophysical compact objects, particularly
black holes. The motion of particles near compact stars and black
holes has been widely reviewed
by~\citet{poisson,laem1,laem2,laem3,zden1,zden2}. Recently, motion
of a particle around the black hole~\citep{rs99,rs98,garcia},
gravitational lensing by black holes~\citep{bozza,wald,gergely},
extraction of energy from black holes~\citep{penrose,dadhich} and
such other effects have been intensively studied. Innermost stable
orbits ~\citep{hartle,barak} as well as scattering and capture
cross section~\citep{zccs,Mizner} of particles in gravitational
field of black holes can be plausible examples for the events
mentioned.

There are no intrinsic electromagnetic fields around the black
holes except the gravitational field~\citep{go64}. In spite of
this black hole can be considered in an exterior asymptotically
uniform magnetic field~\citep{w74,rs99,rs98,Preti}. Under these
assumptions, one can say that the formation of a highly charged
black holes is incredibly in astrophysical conditions~\citep{FN}.
There are several observations showing that there are various
scenarios where the magnetic fields and general relativity can not
be neglected. One of them is the presence of strong magnetic
fields in active galactic nuclei. Another scenario is the
production of relativistic collimated jets in the inner regions of
accretion discs, which can be explained considering
magneto-centrifugal mechanisms. The interplay between
gravitational and electromagnetic interaction is essential for
characteristics of the motion, namely its stability properties.
Motivation for these studies arises from the problem of motion and
acceleration of test particles. The study of the interaction
between particles and electromagnetic fields in curved spacetimes
is also of astrophysical interest, such is the case of strong
synchrotron radiation emerging galactic cores, which can be
explained admitting the existence in those regions of extended and
very intense magnetic fields, interacting with ultrarelativistic
electrons. Such magnetic fields could originate in the inner part
of an accretion disc around the central black hole.

First attempts of building multidimensional models were proposed
by~\citet{k21} in order to unify electromagnetism with gravity.
Then these ideas found reflection in elegant string theory which
is the subject of extensive research in modern physics and promise
to throw light upon many puzzles of nature. One of the most recent
theories including extra dimensions is the braneworld picture of
the Universe.

The braneworld model was first proposed by \citet{RaSu99} assuming
that our four-dimensional space-time is just a slice of
five-dimensional bulk. According to this model only gravity is the
force which can freely propagate between our space-time and bulk
while other fields are confined to four-dimensional Universe. In
this view it is noteworthy to look for effects of the fifth
dimension on our world in frame of theory of gravity i.e. general
relativity. Possible tools for proving braneworld model should be
found from astrophysical objects, namely, compact objects, for
which effects of general relativity are especially strong. For
example, investigations of cosmological and astrophysical
implications of the braneworld theories have been done
by~\citet{maar00,cs01,lan01,hm03,ger06,kg08,MaMu05}. Review of
braneworld models is given e.g. in \citep{maar04}.

For astrophysical interests, static and spherically symmetric
exterior vacuum solutions of the braneworld models were initially
proposed by \cite{Dadhich} which have the mathematical form of the
Reissner-Nordstr\"{o}m solution, in which a tidal Weyl parameter
$Q^\ast$ plays the role of the electric charge squared of the
general relativistic solution.

Observational possibilities of testing the braneworld black hole
models at an astrophysical scale have been intensively discussed
in the literature during the last several years, for example,
through the gravitational lensing~\citep{pk08}, the motion of test
particles~\citep{aaprd,zdnk}, and the classical tests of general
relativity (perihelion precession, deflection of light, and the
radar echo delay) in the Solar System (see \cite{lobo08}). The
energy flux, the emission spectrum, and accretion efficiency from
the accretion disks around several classes of static and rotating
braneworld black holes have been obtained by \citet{pkh08}. The
complete set of analytical solutions of the geodesic equation of
massive test particles in higher dimensional spacetimes which can
be applied to braneworld models is provided in the recent paper
\cite{Lam08}.

The structure of electromagnetic field of spherically and slowly
rotating magnetized star in a Randall-Sundrum II type braneworld
has been considered by~\citet{fa08,ma11} where Maxwell's equations
for the external magnetic field of the slowly rotating star in the
braneworld are analytically solved in approximation of small
distance from the surface of the star. The braneworld version of
the Schwarzschild's interior solution has been abtained
by~\citet{ovalle}. Plasma magnetosphere surrounding rotating
magnetized neutron star in the braneworld has been studied
by~\citet{metal10}. The relativistic quantum interference effects
in the spacetime of slowly rotating object in braneworld as the
Sagnac effect and phase shift effect of interfering particle in
neutron interferometer are derived by~\citet{mht09}. Recently the
magnetized particle motion around black hole in braneworld have
been considered by one of the authors of this paper \citep{rah10}.

This paper is organized as follow: in section \ref{potential}
equations of motion of magnetized particles in the braneworld
spacetime are formulated. For this we use Hamilton-Jacobi equation
which contains new  terms being proportional to the polarization
tensor~\citep[see, e.g.][]{rs99,Preti} which characterizes
magnetization of particle in the equation of motion. Our main aim
consists of checking of influence of this parameter to capture
cross section of particle by black hole in braneworld. And in
section \ref{sec:emfield} we derive the analytical expression for
capture cross section of magnetized particles by braneworld black
hole. Section \ref{sec:ield} is devoted to the study particles
release from capture by black hole. We conclude our results in
Section \ref{conclusion}.

Throughout, we use a space-like signature $(-,+,+, \\
+)$ and a system of units in which $G = 1 = c$, Greek indices run
from 0 to 3, Latin indices from 1 to 3.

\section{Equation of motion for spinning particle}
\label{potential}

The  braneworld  spacetime metric in the spherical coordinates has
the following form \citep{Dadhich}:
\begin{equation}\label{metric}
ds^2=-A^2dt^2+H^2dr^2+r^2d\theta^2+r^2\sin^2\theta \varphi^2,
\end{equation}
 where
$$A^2=H^{-2}=\left(1-\frac{2M}{r}+\frac{Q^*}{r^2}\right),$$
$Q^*$ is the bulk tidal charge and $M$ is the total mass of the
central black hole immersed in an exterior asymptotically uniform
magnetic field $\mathbf{B}_0$. The polar axis is chosen along the
direction of $B_0$.
The Hamilton - Jacobi equation for the magnetized particles motion
has the following form~\citep [See, e.g.][]{rs99}:
\begin{equation}
\label{Ham-Jac} g^{\mu\nu}\left(\frac{\partial S}{\partial
x^\mu}-qA_{\mu}\right)\left(\frac{\partial S}{\partial
x^\nu}-qA_{\nu}\right)=-m^2+mD^{\mu\nu}F_{\mu\nu}\ ,
\end{equation}
where $q$ and $m$ are charge and mass of particle, $A_\alpha$ is
the potential of the electromagnetic field and $F_{\mu\nu}$ is the
electromagnetic field tensor. Nonvanishing components of the
electromagnetic field tensor $F_{\mu\nu}=\partial_\nu A_\mu-
\partial_\mu A_\nu$ are~\citep{rs99}:
\begin{eqnarray}\label{ten el f}
&& {{F_{r\varphi}=B_0r\sin^2\theta}}, \qquad
{{F_{\theta\varphi}=B_0r^2\sin\theta\cos\theta.}}
\end{eqnarray}
The polarization tensor related to antisymmetric spin tensor
$S^{\mu\nu}$ is:
\begin{eqnarray}\label{polar}
&&
D^{\mu\nu}=\frac{q}{m}S^{\mu\nu}=\eta^{\mu\nu\rho\lambda}u_\rho\mu_\lambda\
,
\end{eqnarray}
where $u^{\mu}$ is the 4-velocity and $\mu^{\lambda}$ is the
magnetic moment 4-vector of the particle,
$\eta^{\mu\nu\rho\lambda}$ is the Levi-Civita tensor.

The electromagnetic field tensor can be expressed through the
components of magnetic $B^\alpha$ and electric $E^\alpha$ fields
as
\begin{eqnarray}\label{decomposition}
&& F_{\mu\nu}=\eta_{\mu\nu\alpha\beta}B^\alpha u^\beta+u_\mu
E_\nu- u_\nu E_\mu.
\end{eqnarray}

Using equations (\ref{ten el f}), (\ref{polar}) and
(\ref{decomposition}) one can factorize a constant quantity out of
the interaction term $D\cdot F$ in the following form:
\begin{eqnarray}\label{polar1}
&& D^{\mu\nu}F_{\mu\nu}=2\mu^{\alpha} B_{\alpha}=2\mu B_{0}\cdot
\eta \ ,
\end{eqnarray}
where $\mu$ is the norm of the magnetic moment. Here $\eta$
characterizes magnetization of particle in the following
form~\citep{rs99}:
\begin{equation}\label{eta}
\eta=\beta
\left(1-\frac{2M}{r}+\frac{Q^*}{r^2}\right)\left(1-\frac{2M}{r}+\frac{Q^*}{r^2}-\Omega^2
r^2\right)^{-\frac{1}{2}},
\end{equation}
where $\beta=2\mu B_0/m$ and $\Omega=d\phi/dt$ is the angular
velocity of particle, which is measured by observer at infinity.

It follows  from the equation (\ref{Ham-Jac}) that for the radial
motion of particles:
\begin{equation}\label{f1}
m^2 r^4 \left(\frac{dr}{d\tau}\right)^2=f(r) \ ,
\end{equation}
where
\begin{eqnarray}\label{f}
f(r)=(E^2+m\eta-m^2)r^4+2M(m^2-m\eta)r^3
\nonumber\\
&&\hspace{-6.4cm}+(m\eta Q^*-Q^*-L_{0}^2)r^2+
2ML_{0}^2r-Q^*L_{0}^2,
\end{eqnarray}
$E$ and $L_0$ are the energy and angular momentum of the test
particle, respectively. Dividing equation (\ref{f}) to $m^2$ one
can easily get the polynomial in the form:
\begin{eqnarray}\label{f1}
f_1=\frac{f(r)}{m^2}=(e^2+\eta-1)r^4+2M(1-\eta)r^3
\nonumber\\
&&\hspace{-5.2cm}+(\eta Q^*-Q^*-L^2)r^2+ 2MLr-Q^*L^2,
\end{eqnarray}
where $e=E/m$ and $L=L_0/m$ is the specific energy and angular
momentum of particle, respectively.

In our case particles  are characterized by magnetic moment $\mu$,
but  particles have no electric charge, that is in equation
(\ref{Ham-Jac}) $q=0$  and $\eta$ characterizes the magnetization
of the particle. 

Considering the polynomial $f(r)$ (for $r\geq
r_+=1+\sqrt{1-Q^*}$),  we can obtain the following three types of
motion in the $r$ coordinate~\citep[see, e.g.][]{zah1,zah2}:

1) the polynomial $f(r)$ has no roots for $r\geq r_+$, the
particle then falls into the black hole;

2) the polynomial $f(r)$ has roots $r_{max}>r_+$, for $\partial
f/\partial r > 0$ and the particle departs to infinity after
approaching  the black hole;

3) the polynomial $f(r)$ has a root and $\partial f/\partial r= 0$
and the particle takes an infinite proper time to approach the
surface $r = const$.

For the relativistic particles ($e\gg1$) we obtain expression
\begin{equation}\label{relat}
f(\rho)=(e^2+\eta)\rho^4+2(1-\eta)\rho^3-({\tilde{Q}}^*+l^2)\rho^2+2l^2
\rho-{\tilde{Q}^*} l^2,
\end{equation}
where $\rho=r/M$, $l=L_0/M$ and ${\tilde{Q}}^*=Q^*/M^2$.

In the slow motion limit  when $E=m$, we get the following
equation:
\begin{equation}\label{slow}
f_1(\rho)=\eta\rho^4+2(1-\eta)\rho^3-({\tilde{Q}}^*+l^2)\rho^2+2l^2
\rho-{\tilde{Q}^*} l^2.
\end{equation}
\section{\label{sec:emfield} Capture cross section for magnetized
relativistic particles of a braneworld black hole}
For relativistic particles we can assume $\eta\simeq1$, that is
the value of spin is large and one can obtain from (\ref{relat})
\begin{eqnarray}\label{radial}
\rho^4+\left(\frac{l^2}{e^2+\eta}\right)\rho^2
+\left(\frac{2l^2}{e^2+\eta}\right)\rho-\frac{\tilde{Q^*}l^2}{e^2+\eta}=0\
,
\end{eqnarray}
since $\tilde{Q}^*\ll l^2$ one can neglect the fourth term in the
left hand side of the equation. It is known that the symmetric
polynomials  (when $n=4$) have the form~\citep[see,
e.g.][]{Kostrikin}
\begin{equation}\label{polinom}
p_k={X_1}^k+{X_2}^k+{X_3}^k+{X_4}^k.
\end{equation}
Expressing polynomials $p_k$ in terms of $s_k$ and using equation
of the Newton according to~\citep[see, e.g.][]{Kostrikin}, we
calculate the polynomials and the  discriminant of the family
$X_k$ in roots of the polynomial $f(\rho)$ as
\begin{eqnarray}\label{p}
p_1=s_1=0,  p_2=-2s_2, p_3=3s_3, p_4=2{s_2}^2-4{s_4}
\nonumber\\
&&\hspace{-8.2cm} p_5=-5s_3s_2,  p_6=-2{s_2}^3+3{s_3}^2+6s_4s_2.
\end{eqnarray}
Using again formula of the Newton \citep[see, e.g.][]{Kostrikin}
for polynomials one  may obtain the following expression
\begin{equation}\label{s}
s_1=0,  s_2=-\frac{l^2}{e^2+\eta},  s_3=-\frac{2l^2}{e^2+\eta},
s_4=-\frac{\tilde{Q^*}l^2}{e^2+\eta}.
\end{equation}
Introducing the notation $\alpha=(e^2+\eta)^{-1}$, we get from
(\ref{s}) the following expressions for the symmetrical
polynomials
\begin{eqnarray}\label{f1}
p_1=0, \quad p_2=2l^2\alpha, \quad p_3=-6l^2\alpha,
\nonumber\\
&&\hspace{-6.1cm}p_4=2l^2\alpha(l^2\alpha+2\tilde{Q^*}),
\nonumber\\
&&\hspace{-6.1cm}p_5=-10l^2\alpha, \quad
p_6=2l^2\alpha(l^2\alpha+3\tilde{Q^*}+6).
\end{eqnarray}
Calculating the determinant for symmetrical polynomials
\citep[see, e.g.][]{Kostrikin} we get equation
\begin{equation}\label{l1}
\ell^2(1-\tilde{Q^*})+\ell(36\tilde{Q^*}-8\tilde{Q^*}^2-27)-16\tilde{Q^*}^3=0,
\end{equation}
where $\ell=l^2/(e^2+\eta)$. If $\tilde{Q^*}=0$ and $\eta=0$, we
get for the classical particle (around the Schwarzschild black
hole) the well known expression $l^2=27$ or $L_{cr}=3\sqrt{3}$.

For the spinning particles we have from (\ref{l1}):
\begin{equation}\label{l2}
\frac{l^2}{e^2+\eta}=27  \quad for  \quad \tilde{Q^*}=0,
\end{equation}
and
\begin{equation}\label{l2}
\frac{l^2}{e^2+\eta}=16  \quad for  \quad \tilde{Q^*}=1.
\end{equation}

And solving the quadratic equation (\ref{l1}) we get (we neglect
the term being proportional to $\tilde{Q}^2$):
\begin{equation}\label{l3}
\frac{l^2}{e^2+\eta}=\frac{27-36\tilde{Q^*}+27(1-2,66\tilde{Q^*})^{\frac{1}{2}}}{2(1-\tilde{Q^*})}.
\end{equation}
From the equation (\ref{l3}) one may obtain the known expression
for the classical particle when $\eta=0$. One may see that when
$e/(l+\eta)$ is constant the particles can move along the
classical orbits with the smaller energy than the particles with
zero spin ($\eta=0$).
\section{\label{sec:ield} Release of particles from black hole in braneworld}
\begin{figure*}
a) \includegraphics[width=0.45\textwidth]{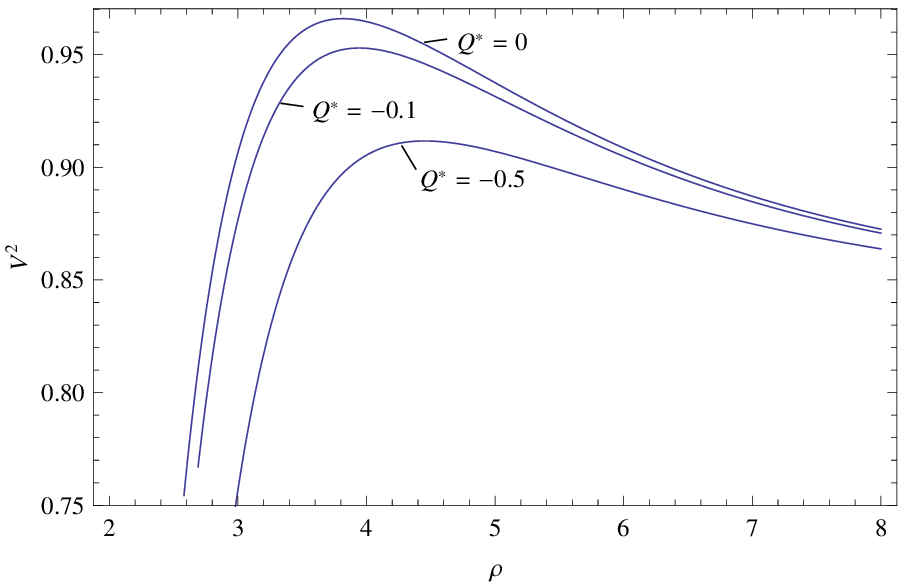}
b) \includegraphics[width=0.45\textwidth]{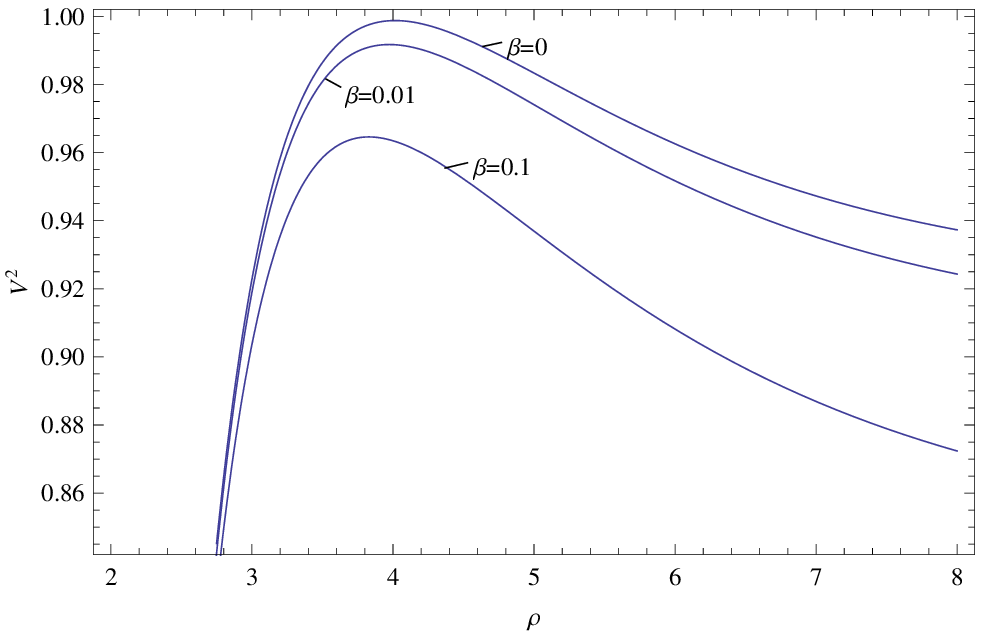}
\caption{\label{effpot} The radial dependence of the effective
potential of the radial motion of the magnetized particle around
black hole in braneworld for the different values of the (a)
dimensionless parameter $Q^*$ and (b) dimensionless magnetic
parameter $\beta$.}
\end{figure*}

We will consider now particles release from capture by black hole,
that is, we  will  write the expression for the critical values of
impact parameter and angle of deflection of particles  trajectory.
Let us rewrite the equation for radial motion of test magnetized
particles around black hole in braneworld in external
asymptotically uniform magnetic field in the following form:
\begin{eqnarray}
\left(\frac{dr}{d\tau}\right)^2=e^2-V^2\ ,
\end{eqnarray}
where:
\begin{eqnarray}
V^2
&=&\left[\frac{1}{\rho^2}-\beta\left(1-\frac{2}{\rho}+\frac{\tilde{Q^*}}{\rho^2}\right)^{{1}/{2}}
\right]\nonumber\\
&&\times\left(1-\frac{2}{\rho}+\frac{\tilde{Q^*}}{\rho^2}\right)
\label{V}
\end{eqnarray}
is the effective potential of radial motion of magnetized test
particle for rest observer($\Omega=0$) in the limiting case when
impact parameter $b=l/e$ and neglecting small terms of ${\cal
O}(1/L^2)$ . In Fig. \ref{effpot} the radial dependence of the
effective potential for radial motion of magnetized particle
around black hole in braneworld immersed in external
asymptotically uniform magnetic field for the different values of
the dimensionless brane parameter $\tilde{Q^{*}}$ and magnetic
parameter $\beta$ are shown.  It is easy to see that orbits of the
particles become more stable with increasing of the parameter
$\beta$. As it is seen from the figure the particle coming from
infinity and passing by the source will be not captured: it will
be reflected and go to the infinity again. The orbits start to be
only parabolic or hyperbolic and no more circular or elliptical
orbits exist with increasing the dimensionless parameter $\beta$,
i.e. captured magnetized particles by the central object are going
to leave the black hole in braneworld. It should be noted that the
influence of the magnetic parameter to the trajectory of the
magnetized particles is not sufficient near the black hole, it
comes to be more sufficient at far distances from the compact
object.

It it is well known that the  classical solution for the effective
potential has maximum value at $\rho=3$  and one can see that
$b_{cr}=3\sqrt{3}$, whereas for the spinning particles we obtain
\begin{equation}\label{b}
b_{cr}=\frac{3\sqrt{3}}{\sqrt{(1-\tilde{\eta})(1+3\tilde{Q^*})}},
\end{equation}
where $\tilde{\eta}=M^2\eta/l^2$, $\eta$ is the term used above
and $l^2/M^2$ is the dimensionless angular momentum.

The proper observer in this coordinate system can calculate the
velocity of particle in the following way~\citep[see,
e.g.][]{Mizner}: $\vartheta_r=\pm\sqrt{1-b^2/B^2}$ and
$\vartheta_{\phi}=b/B$, where
$B=\sqrt{(1-2/\rho+\tilde{Q^*}/\rho^2)(1/\rho^2-\tilde{\eta})}$
and from $\theta_{\rm
cr}=\arccos\vartheta_r=\arcsin\vartheta_{\phi}$ we can find the
angle $\theta_{\rm cr}$, which between the direction of
propagation and radial direction of particles motion
\begin{eqnarray}\label{angle}
\theta_{\rm cr}
&=&\arcsin\Bigg\{3\Bigg[3\left(1-\frac{2}{\rho}+\frac{\tilde{Q^*}}{\rho^2}\right)
\nonumber\\
&& \times\left(\frac{1}{\rho^2}
-\beta\sqrt{1-\frac{2}{\rho}+\frac{\tilde{Q^*}}{\rho^2}}\right)\Bigg]^{1/2}
 \Bigg\}.
\end{eqnarray}

\begin{figure}
\includegraphics[width=0.45\textwidth]{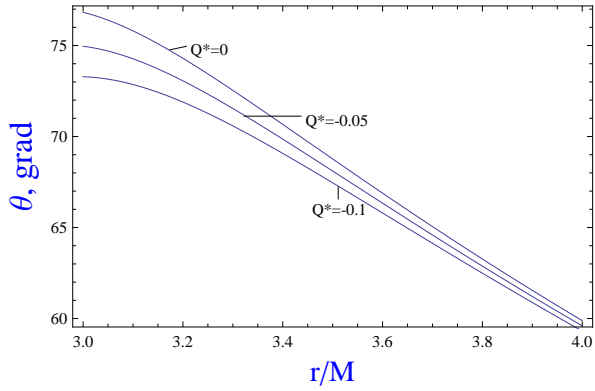}
\caption{\label{fig_a} The radial dependence of the angle of
release for the different values of the brane parameter.}
\end{figure}

In the Fig.\ref{fig_a} the radial dependence of angle of release
for the different values of the brane parameter is shown. One can
see from the dependence that the angle of capture is decreased
with the increase of the module of brane parameter.

\begin{figure}
\includegraphics[width=0.45\textwidth]{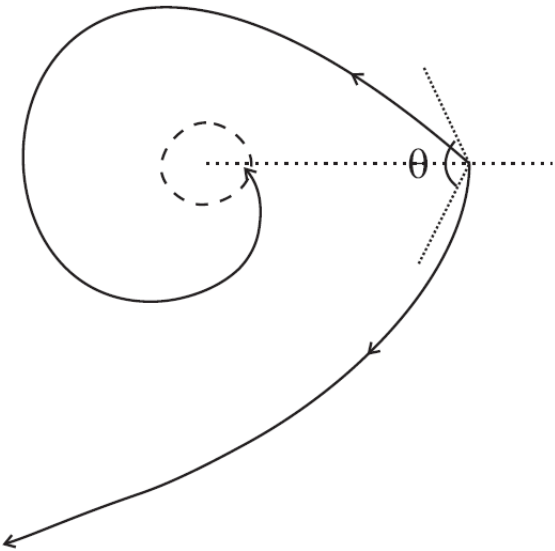} %

\caption{\label{scheme} The particles trajectories. If initially
particle's motion direction is lower than $\theta$, particle will
be captured by the central black hole, otherwise particles can
escape from the black hole. Dashed circle represents the event
horizon of the black hole.}
\end{figure}

The physical meaning of the critical angle $\theta_{\rm cr}$ can
be explained using the scheme presented in Fig.\ref{scheme}.
Particles directed with the angle lower than $\theta_{\rm cr}$
will be captured by the central black hole, otherwise particles
can escape from the central black hole.


\section{\label{conclusion} Conclusion}

Here we have derived analytical expressions for capture cross
section of magnetized particles by  black hole in braneworld
immersed in external magnetic field. The expressions for capture
cross sections were obtained using the Hamilton-Jacobi formalism.
The form of Hamilton-Jacobi equation was chosen as
in~\citep{rs99,preti1}. Such analysis was first performed by
Zakharov in Ref. \cite{zccs} for particles with zero magnetic
moment. Extensive analysis of the effective potential of the
radial motion for the magnetized test particles around black hole
in braneworld has shown that the orbits of the particles become
more stable with increasing of the parameter $\beta$. It was shown
that the particle coming from infinity and passing by the source
will be not captured: it will be reflected and go to the infinity
again. Captured magnetized particles by the central object are
going to leave the black hole in braneworld. It should be noted
that the influence of the magnetic parameter to the trajectory of
the magnetized particles is not sufficient near the black hole, it
comes to be more sufficient at far distances from the compact
object. The influence of the brane parameter is sufficient near
the object and brane parameter forces stable circular orbits to
shift to the observer at the infinity. The similar result has been
obtained by~\citet{aaprd} and they have obtained upper limit for
the dimensionless brane parameter comparing the theoretical
results with the astrophysical data.

One of the main interesting tasks is to find the lower limit for
the critical angular momentum of magnetized particles. Particles
with angular momentum lower than the critical value will be
captured by the  black hole in braneworld. It is well known that
for the neutral particles the critical value of angular momentum
is $L_{\rm cr}=4M$. In this paper we have shown that in the
presence of the brane parameter, the external magnetic field and
magnetic moment of the particle this critical value is decreased,
i.e. magnetized particle to be captured can escape from the
central black hole in braneworld immersed in external magnetic
field.

In this paper the angle of direction of particle's motion, that
is, in what direction particles can release from the capture of
black hole in braneworld have been also considered. We have found
the radial dependence of critical angle $\theta_{\rm cr}$ for
particles for the different values of the parameter $Q^{*}$.
Particles directed with angle lower than $\theta_{\rm cr}$ will be
captured by the central black hole in braneworld. In this paper we
have concluded that in the presence of nonvanishing brane
parameter $Q^{*}$ the critical value of angle decreases. Particle
with the magnetic moment are more stable with compare to the
neutral particles around braneworld black hole immersed in
external magnetic field.

\section*{\label{acknow} Acknowledgments}

This research is supported in part by the UzFFR (projects 1-10 and
11-10) and projects FA-F2-F079 and FA-F2-F061 of the UzAS.


\begin{thebibliography}{99}

\bibitem[\protect\citeauthoryear{{Abdujabbarov et. al}}{2010}]{aaprd}
{Abdujabbarov A. A., Ahmedov B. J.: Phys. Rev. D \textbf{81},
044022 (2010)}

\bibitem[\protect\citeauthoryear{{Barak and Sago}}{2009}]{barak}
Barak, L., Sago, N.: {Phys. Rev. D} {\bf 102}, 191101 (2009)

\bibitem[\protect\citeauthoryear{{B\"{o}hmer et al.}}{2008}]{lobo08}
{B\"{o}hmer, C.G., Harko, T., Lobo, F.S.N.: Class. Quantum Grav.
{\bf 25}, 045015 (2008)}

\bibitem[\protect\citeauthoryear{{Bozza}}{2010}]{bozza}
Bozza, V.: {Gen. Rel. Grav.} {\bf{42}}, 2269 (2010)


\bibitem[\protect\citeauthoryear{{Campos and Sopuerta}}{2001}]{cs01}
Campos, A., Sopuerta, C.F.: Phys. Rev. D {\bf 63}, 104012 (2001)

\bibitem[\protect\citeauthoryear{Chandrasekhar}{1983}]{Chandra}
Chandrasekhar S.: \textit{{The mathematical theory of black
holes}}, (Oxford Univ. Press, 1983)

\bibitem[\protect\citeauthoryear{{Dadhich et al.}}{2000}]{Dadhich}
{Dadhich N., Maartens, R., Papodopoulos,  P.,  Rezania, V.: Phys.
Lett. B {\bf 487}, 1 (2000)}

\bibitem[\protect\citeauthoryear{{Enolski et al.}}{2011}]{laem1}
{Enolski, V.Z., Hackmann, E., Kagramanova, V., Kunz, J.,
L\"{a}mmerzahl, C.:  {J. Geom. Phys.} \textbf{61}, 899 (2011)}

\bibitem[\protect\citeauthoryear{Fattoev and Ahmedov}{2008}]{fa08}
Fattoev, F.J., Ahmedov, B.J.: Phys. Rev. D {\bf 78}, 047501 (2008)

\bibitem[\protect\citeauthoryear{{de Felice} and {Sorge}}{2003}]{rs99}
{de Felice, F. Sorge, F.: Class. Quantum Gravit. \textbf{20}, 469
(2003)}

\bibitem[\protect\citeauthoryear{{de Felice et al.}}{2004}]{rs98}
{de Felice, F., Sorge, F., Zilio, S.: Class. Quantum Gravit. {\bf
21}, 961 (2004)}

\bibitem[\protect\citeauthoryear{{Garcia-Duque and Garcia-Reyes}}{2010}]{garcia}
Garcia-Duque, C. H. Garcia-Reyes, G.:  arXiv:1009.1084v5
[gr-qc](2101)

\bibitem[\protect\citeauthoryear{{Gergely and Dar\'{a}zs}}{2006}]{gergely}
Gergely, L., Dar\'{a}zs, B.: {Publications of the Astronomy
Department of the E$\ddot{o}$tv$\ddot{o}$s University } {\bf{17}},
213 (2006)

\bibitem[\protect\citeauthoryear{{Gergely}}{2006}]{ger06}
Gergely, L.A.: Phys. Rev. D {\bf 74}, 024002 (2006)

%
\bibitem[\protect\citeauthoryear{{Ginzburg and Ozernoy}}{1964}]{go64}
      {Ginzburg, V.L., Ozernoy L.M.: Zh. Eksp. Teor. Fiz.
      {\textbf{47}}, 1030 (1964) [English
      version: Sov.Phys. JETP \textbf{20}, 689 (1964)]}

\bibitem[\protect\citeauthoryear{{Hackman et al.}}{2008}]{Lam08}
{Hackman, E., Kagramanova, V., Kunz, J., L\"{a}mmerzahl, C.: Phys.
Rev. D {\bf 78}, 124018 (2008)}

\bibitem[\protect\citeauthoryear{{Hackmann et al.}}{2011}]{laem3}
{Hackmann, E., Hartmann, B., L\"{a}mmerzahl, C., Sirimachan, P.:
{Phys. Rev. D} \textbf{82}, 044024 (2011)}

\bibitem[\protect\citeauthoryear{{Harko and Mak}}{2003}]{hm03}
Harko, T., Mak, M.K.: Class. Quantum Grav. {\bf 20}, 407 (2003)

\bibitem[\protect\citeauthoryear{{Hartle}}{2003}]{hartle}
Hartle, J.B.: \textit{Gravity: an introduction to Einstein's
general relativity} (San Francisco: Addison Wesley) (2003)

\bibitem[\protect\citeauthoryear{{Hartmann et al.}}{2011}]{laem2}
{Hartmann, B., L\"{a}mmerzahl, C., Sirimachan, P.: {Phys. Rev. D}
\textbf{83}, 045027 (2011)}

\bibitem[\protect\citeauthoryear{{Holz and Wald}}{1998}]{wald}
Holz, D., Wald, R.;  {Phys. Rev. D} {\bf{58}}, 063501 (1998)

\bibitem[\protect\citeauthoryear{{Kaluza}}{1921}]{k21}
Kaluza, T.:  On The Problem Of Unity In Physics. Sitzungsber.
Preuss. Akad. Wiss. Berlin (Math. Phys.), 966-972 (1921)

\bibitem[\protect\citeauthoryear{Kostrikin}{1994}]{Kostrikin}
Kostrikin A. I.   {\it{Introduction to algebra}}, (Moskow: Nauka,
1994)

\bibitem[\protect\citeauthoryear{{Kovacs and Gergely}}{2008}]{kg08}
Kovacs, Z., Gergely, L.A.:  Phys. Rev. D {\bf 77}, 024003 (2008)

\bibitem[\protect\citeauthoryear{{Kovar et al.}}{2010}]{zden1}
{Kovar, J., Kopacek, O., Karas, V., Stuchlik, Z.: {Class. Quantum
Grav.} \textbf{27}, 135006 (2010)}

\bibitem[\protect\citeauthoryear{{Langlois}}{2001}]{lan01}
Langlois, D.: Phys. Rev. Lett. {\bf 86}, 2212 (2001)

\bibitem[\protect\citeauthoryear{{Maartens}}{2000}]{maar00}
Maartens, R.: Phys. Rev. D {\bf 62}, 084023 (2000)

\bibitem[\protect\citeauthoryear{{Maartens}}{2004}]{maar04}
Maartens, R.: Living Reviews in Relativity {\bf  7}, 1 (2004)

\bibitem[\protect\citeauthoryear{{Majumdar and Mukherjee}}{2005}]{MaMu05}
Majumdar, A.S., Mukherjee, N.: Int. J. Mod. Phys. D {\bf 14}, 1095
(2005)

\bibitem[\protect\citeauthoryear{Mamadjanov et al.}{2010}]{mht09}
Mamadjanov, A.I., Hakimov, A.A., Tojiev, S.R.: Mod. Phys. Lett. A
{\bf 25}, 243 (2010)

\bibitem[\protect\citeauthoryear{Misner et al.}{1973}]{Mizner}
 Misner C. W.,  Thorne K.S. and  Wheeler J.A.: {\textit
 {Gravitation}},
(San Francisco: Freeman, 1973)

\bibitem[\protect\citeauthoryear{Morozova and Ahmedov}{2011}]{ma11}
Morozova, V. S., Ahmedov, B. J.: Astrophys. Space Sci. {\bf 333},
133 (2011)

\bibitem[\protect\citeauthoryear{Morozova et al.}{2010}]{metal10}
Morozova, V. S., Ahmedov, B. J., Abdujabbarov, A.A., Mamadjanov
A.I.: Astrophys. Space Sci. {\bf  330}, 257 (2010)

\bibitem[\protect\citeauthoryear{Novikov et al.}{1998}]{FN}
{ Novikov I.D. and  Frolov V.P. {\textit {Black Hole Physics}},  (
Dordrecht: Kluwer Academic, 1998)}

\bibitem[\protect\citeauthoryear{Ovalle}{2010}]{ovalle}
     Ovalle, J.: Mod. Phys. Lett. A, {\bf 25}, 3323 (2010)

\bibitem[\protect\citeauthoryear{Pal and Kar}{2008}]{pk08}
    Pal, S., Kar, S.: Class. Quantum Grav., {\bf 25}, 045003
    (2008)

\bibitem[\protect\citeauthoryear{{Penrose}}{1969}]{penrose}
{Penrose, R.: {J. Math. Phys.} \textbf{10}, 38 (1969)}

\bibitem[\protect\citeauthoryear{{Poisson}}{2004}]{poisson}
{Poisson, E.: {Living Rev. Relativity} {\bf 7} 6 URL:
\\http://www.livingreviews.org/lrr-2004-6  (2004)}


\bibitem[\protect\citeauthoryear{{Prabhu and Dadhich}}{2010}]{dadhich}
{Prabhu, K., Dadhich, N.: {Phys. Rev. D} \textbf{81}, 024011
(2010)}

\bibitem[\protect\citeauthoryear{Preti}{2004a}]{Preti}
Preti, G.: Class. Quantum Gravit. {\bf 21}, 3433 (2004a)

\bibitem[\protect\citeauthoryear{Preti}{2004b}]{preti1}
Preti, G.: {Phys. Rev. D} {\bf 70}, 024012 (2004b)

\bibitem[\protect\citeauthoryear{{Pun et al.}}{2008}]{pkh08}
{Pun, C.S.J., Kov\'{a}cs, Z., Harko, T.: Phys. Rev. D {\bf 78},
084015 (2008)}

\bibitem[\protect\citeauthoryear{Rahimov}{2011}]{rah10}
Rahimov, O.G.: Mod. Phys. Lett. A, {\bf 26}, 399 (2011)

\bibitem[\protect\citeauthoryear{{Randall and Sundrum}}{1999}]{RaSu99}
Randall, L., Sundrum, R.: Phys. Rev. Lett. {\bf 83}, 3370 (1999)

\bibitem[\protect\citeauthoryear{{Schee and Stuchlik}}{2009}]{zdnk}
    Schee, J., Stuchlik, Z.: Gen. Rel. Gravit., {\bf 41}, 1795 (2009)

\bibitem[\protect\citeauthoryear{Shapiro et al.}{1983}]{Shapiro}
Shapiro S. and Teukolsky S.:  {\textit { Black holes, White Dwarfs
and Neutron Stars}}, (New York: Wiley,  1983)

\bibitem[\protect\citeauthoryear{{Stuchlik and Hledik}}{2000}]{zden2}
{Stuchlik, Z., Hledik, S.: {Class. Quantum Grav.} \textbf{17},
4541 (2000)}

%
\bibitem[\protect\citeauthoryear{{Wald}}{1974}]{w74}
       {Wald, R.M.: Phys. Rev. D \textbf{10}, 1680 (1974)}

\bibitem[\protect\citeauthoryear{Zakharov}{1986}]{zah1}
Zakharov, A.F.: Sov. Phys-JETP. {\bf 64}, 1 (1986)

\bibitem[\protect\citeauthoryear{Zakharov}{1991}]{zah2}
Zakharov, A.F.: ITEP. Preprint {\bf 44}, 1 (1991)

\bibitem[\protect\citeauthoryear{Zakharov}{1994}]{zccs}
Zakharov, A.F.: Class. Quantum Gravit. {\bf 11}, 1027 (1994)

\bibitem[\protect\citeauthoryear{Zeldovich et al.}{1971}]{Zeldovich}
Zeldovich, Y. B., and  Novicow, I. D.: {\textit {Theory of gravity
and evolution of stars}}, (Moscow: Nauka, 1971)



\end{thebibliography}
\end{document}